\documentclass[final,5p,times,twocolumn]{elsarticle}

\usepackage{amssymb}
\usepackage{lipsum}
\usepackage{amsmath}
\usepackage{graphicx}
\usepackage[colorlinks=true, allcolors=blue]{hyperref}
\usepackage{esvect}
\usepackage{cancel}
\usepackage{physics}
\usepackage{float}
\usepackage{mathtools}
\usepackage{multirow}
\usepackage{braket}
\usepackage{gensymb}
\usepackage{orcidlink}

\journal{Physics Letters B}

\begin{document}

\begin{frontmatter}

\title{Can Coulomb breakup of $N=Z$ clustered nuclei constrain capture cross sections? \\
Analyzing the $\alpha(d,\gamma)\prescript{6}{}{\rm Li}$ case.}

\author[1]{M. Sanjinez-Ortiz\,\orcidlink{0009-0000-8957-8885}}
\ead{msanjine@uni-mainz.de}

\author[1]{P. Capel\,\orcidlink{0000-0001-8722-7133}}
\ead{pcapel@uni-mainz.de}

\affiliation[1]{organization={Institut für Kernphysik, Johannes Gutenberg Universität-Mainz},
            city={Mainz},
            postcode={D-55099}, 
            country={Germany}}

\begin{abstract}
In nuclear astrophysics, it is fundamental to measure the cross section of radiative-capture reactions involving charged particles at low energy. However, this endeavor is difficult. Coulomb breakup of nuclei on heavy targets has been proposed as an indirect method to infer the low-energy cross sections, because this reaction corresponds to the time reversed process of radiative capture. Two Coulomb breakup experiments aiming to infer the $\alpha(d, \gamma)\prescript{6}{}{\rm Li}$ radiative-capture cross section have been performed at 26 MeV/nucleon and 150 MeV/nucleon. Within a fully dynamical reaction model, we report here on a new theoretical analysis of these reactions. Our analysis indicates that the breakup of $\prescript{6}{}{\rm Li}$ onto $\prescript{208}{}{\rm Pb}$ is nuclear dominated at forward angles because Coulomb breakup is suppressed due to the $N=Z$ cluster structure of $\prescript{6}{}{\rm Li}$. Our analysis also suggests that the reaction is characterized by marked Coulomb-nuclear interferences. Moreover, we find that higher-order effects considerably increase the breakup cross section at low energies. Consequently, extracting radiative capture cross sections from the breakup reaction data in this case is extremely challenging. We suggest that the forward Coulomb breakup suppression is a general feature of collision processes involving $N=Z$ clustered nuclei. Our findings underscore the importance of advancing alternative indirect methods for nuclear reactions in astrophysics such as electron-induced dissociation.
\end{abstract}


\end{frontmatter}

\section{Introduction}
\label{introduction}

Indirect methods for the determination of reaction cross sections relevant in astrophysics are important due to the experimental limitations in performing direct measurements at the energies of interest. In particular, the breakup or dissociation of nuclear projectiles under the Coulomb field of heavy targets has been presented as an interesting indirect method having the potential to yield information on reactions involving electromagnetic processes such as radiative captures \cite{Baur}. This method, often referred to as \textit{Coulomb breakup}, has been used experimentally in different occasions to infer the cross section of capture reactions involving both neutral and charged fragments \cite{Baur:1996df}.

A radiative capture of particular interest in astrophysics is the $\alpha(d,\gamma)\prescript{6}{}{\rm Li}$ reaction, which plays a relevant role in the picture of Big-Bang nucleosynthesis (BBN). Being the main path for the production of primordial $\prescript{6}{}{\rm Li}$, determining the reaction rate for this radiative process has been of utmost importance to interpret observationally determined abundances of lithium isotopes in old stars \cite{ asplund2006lithium, Spite_2009}, and to understand the production of other light nuclei in the early universe \cite{Wiescher}.

The cross section of this reaction was measured directly at energies corresponding to the $3^+$ resonance and beyond by \citet{robertson1981observation} and \citet{mohr1994direct}. However, the determination in the BBN energy range remained prohibitive at the time. Thus, the Coulomb breakup method was used in two occasions in attempts to determine the $\alpha(d,\gamma)\prescript{6}{}{\rm Li}$ low-energy cross section, both using $\prescript{208}{}{\rm Pb}$ as a heavy target.

The first experimental attempt was performed at the Karlsruhe Isochronous Cyclotron by \citet{Kiener} with a $\prescript{6}{}{\rm Li}$ beam at 26 MeV/nucleon, and the second one at the GSI in Darmstadt by \citet{Hammache} with a 150 MeV/nucleon beam. Only the analysis by \citet{Kiener} led to a concrete determination of the reaction S-factor in the energy range of interest for BBN, however, their results have been noted to be in tension with several theoretical predictions \cite{Hammache, hebborn}. 

Later, the LUNA collaboration obtained the first direct measurement of the $\alpha(d,\gamma)\prescript{6}{}{\rm Li}$ cross section in the energy range of BBN interest \cite{LUNA}. The disagreement between these direct measurements and the indirect measurements by \citet{Kiener} has been noticed and has raised discussions. In this context, nuclear processes polluting the indirect measurements  have been mentioned as a possible source of the disagreement, among other possibilities \cite{Hammache, hebborn, igamov2000triple}.

In this Letter, we present a new theoretical analysis of the Coulomb breakup reactions measured by \citet{Kiener} and \citet{Hammache}. This analysis is performed to gain better insights into the discrepancy between direct and indirect measurements of the $\alpha(d,\gamma)\prescript{6}{}{\rm Li}$ reaction cross section as well as into the dynamics of Coulomb breakup, in particular for $N=Z$ clustered nuclei.

The structure of the paper is as follows. In Sec. 2 we introduce the most important aspects of the breakup reaction model used for the analysis. In Sec. 3, we present our results at the two considered beam energies. Section 4 comments on the inference of the astrophysical S-factor from breakup reaction measurements under the assumptions of \citet{Kiener}. Finally, Sec. 5 presents a summary of our findings and conclusions.

\section{Breakup-reaction model}

The analysis of the breakup reactions in this work was done within the Dynamical Eikonal Approximation (DEA) \cite{Baye}. The DEA is a reaction framework that combines the advantages of semiclassical and eikonal models by including dynamical and quantum effects. It has been extensively used to model the scattering and dissociation of weakly bound nuclei at different experimental conditions \cite{Goldstein, Goldstein2, Esbensen}. 

We consider the breakup of a projectile $ P$ with mass $m_{ P}$ due to the interaction with a target nucleus $T$ with mass $m_T$. The projectile consists of two structureless fragments, an $\alpha$ particle and a deuteron in our case. Considering \textbf{\textit{r}} to be the relative coordinate between those fragments, the properties of the projectile are described by its internal Hamiltonian

\begin{equation}
    H_0 = -\frac{\hbar^2}{2 \mu_{P}} \Delta_\textbf{\textit{r}} + V_{\alpha d}(\textbf{\textit{r}}),
\end{equation}

\noindent
where $\mu_{P}$ is the $\alpha$-$d$ reduced mass, and $V_{\alpha d}$ corresponds to an effective potential modeling the inter-cluster interaction. The Hamiltonian of the three-body system including the target reads, after removal of the center-of-mass motion \cite{Baye}

\begin{equation}
    H = -\frac{\hbar^2}{2 \mu} \Delta_\textbf{\textit{R}} + H_0 + V_{\alpha T} \left( \textbf{\textit{R}} - \frac{m_d}{m_P} \textbf{\textit{r}}\right) + V_{dT}\left( \textbf{\textit{R}} + \frac{m_\alpha}{m_P} \textbf{\textit{r}}\right) ,
\end{equation}

\noindent
where $\mu$ is the projectile-target reduced mass, $\textbf{\textit{R}}$ is the coordinate of the center-of-mass of the projectile relative to the target. Here, $V_{\alpha T}$ and $V_{dT}$ are optical potentials that simulate the interactions between the projectile fragments and the target.

We solve the Schrödinger equation $H \Psi(\textbf{\textit{R}}, \textbf{\textit{r}}) = E_t \Psi(\textbf{\textit{R}}, \textbf{\textit{r}})$ with the initial condition ${\Psi} \xrightarrow[Z \to - \infty]{} e^{iKZ} \phi_0 $, where \textit{K} is the wave number related to the total energy as

\begin{equation}
    E_t = \frac{h^2K^2}{2\mu} + E_0 , 
\end{equation}

\noindent
with $E_0$ the energy corresponding to the ground-state wave function $\phi_0$ of the projectile. At the intermediate beam energies considered here, we
introduce the eikonal ansatz \cite{Glauber}

\begin{equation}
    \Psi(\textbf{\textit{R}}, \textbf{\textit{r}}) = e^{iKZ}\hat{\Psi}(\textbf{\textit{R}}, \textbf{\textit{r}}). 
    \label{eq:eikonal_g}
\end{equation}

\noindent
With this factorization, the Schrödinger equation then becomes

\begin{equation}
    \frac{\hbar^2}{2\mu} \Delta_\textbf{\textit{R}} \hat{\Psi} + i\hbar v \frac{\partial \hat{\Psi} }{\partial Z} =  (H_0 + V_{\alpha T} + V_{dT} - E_0) \hat{\Psi},
    \label{eq:eikonal}
\end{equation}

\noindent
where $v = \hbar K/\mu$ is the asymptotic $P$-$T$ relative velocity. The eikonal approximation within the DEA consists in neglecting the first term in Eq. (\ref{eq:eikonal}), and the eikonal solution is obtained by solving

\begin{equation}
    i\hbar v \frac{\partial \hat{\Psi}}{\partial Z} = (H_0 + V_{\alpha T} + V_{dT} - E_0) \hat{\Psi},
    \label{eq:eikonal2}
\end{equation}

\begin{figure}[!t]
\begin{center}
    \includegraphics[width=0.92\linewidth]{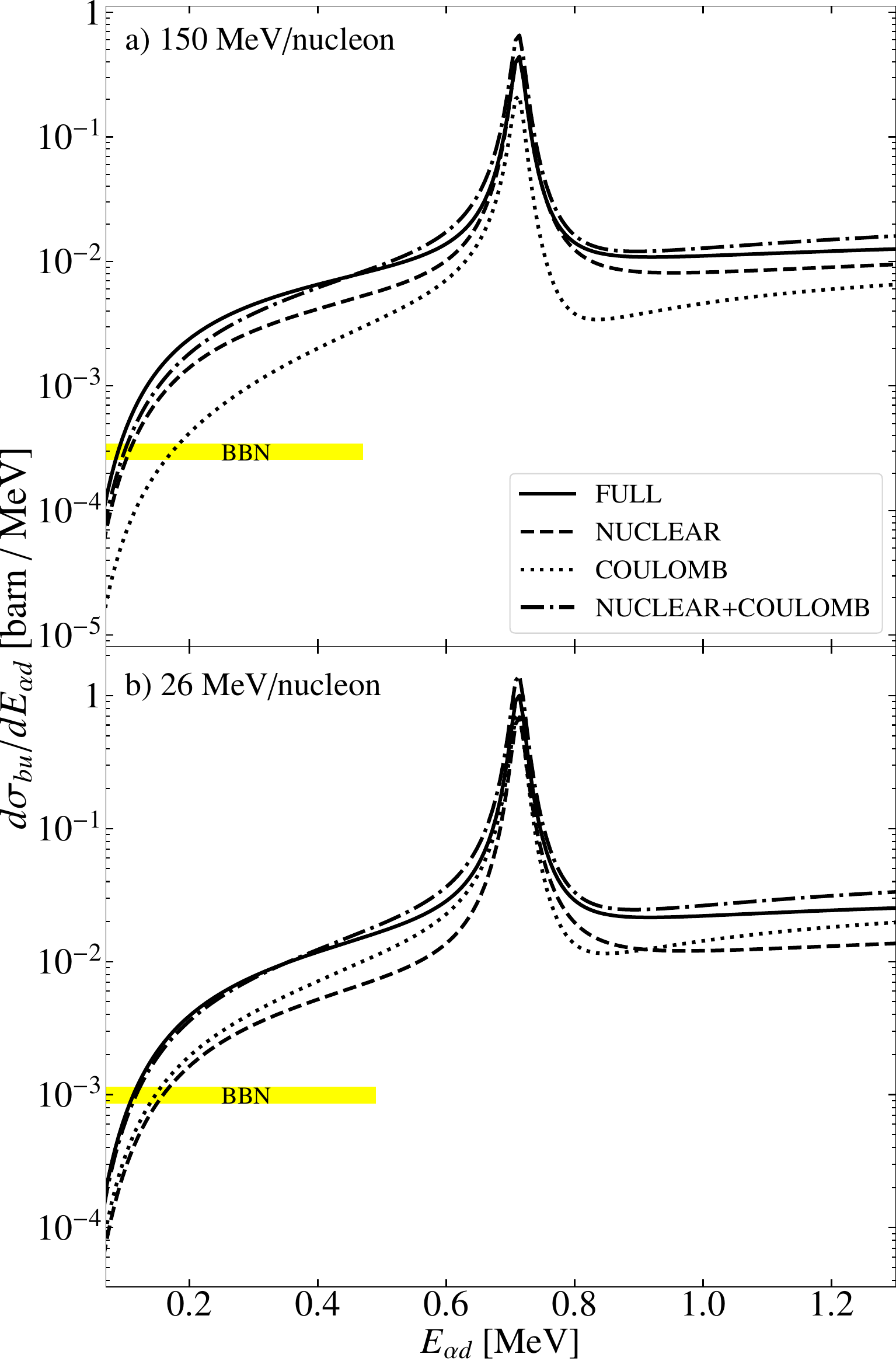}
    \caption{Cross section for the breakup of $\prescript{6}{}{\rm Li}$ onto $\prescript{208}{}{\rm Pb}$ as a function of the final  $\alpha$-$d$ relative energy $E_{\alpha d}$ at a) 150 MeV/nucleon and b) 26 MeV/nucleon. The solid line (---) corresponds to the full breakup cross section, while the dotted ($\cdots$) and dashed ($--$) lines correspond to pure Coulomb and pure nuclear breakup calculations, respectively. The dash-dotted ($- \cdot -$) line is the incoherent sum of pure Coulomb and pure nuclear calculations.}
     \label{fig:energy}
\end{center}
\end{figure}

\noindent
with the initial condition $\hat{\Psi} \xrightarrow[Z \to - \infty]{} \phi_0$. The DEA Eq. (\ref{eq:eikonal2}) is equivalent in form to the semiclassical time-dependent Schrödigner equation for straight-line trajectories and can be solved with the same numerical approach, as described in detail in Ref. \cite{Capel3}. 

With the approximate eikonal solution, we then compute the different transition matrix elements for the dissociation process, see Refs. \cite{Baye, Goldstein}. From these, we determine the dissociation cross sections as a function of the $\alpha$-$d$ relative energy $E_{\alpha d}$ and as a function of the scattering angle $\theta$ of their center-of-mass.

\begin{figure*}
    \centering
    \includegraphics[width=0.92\textwidth]{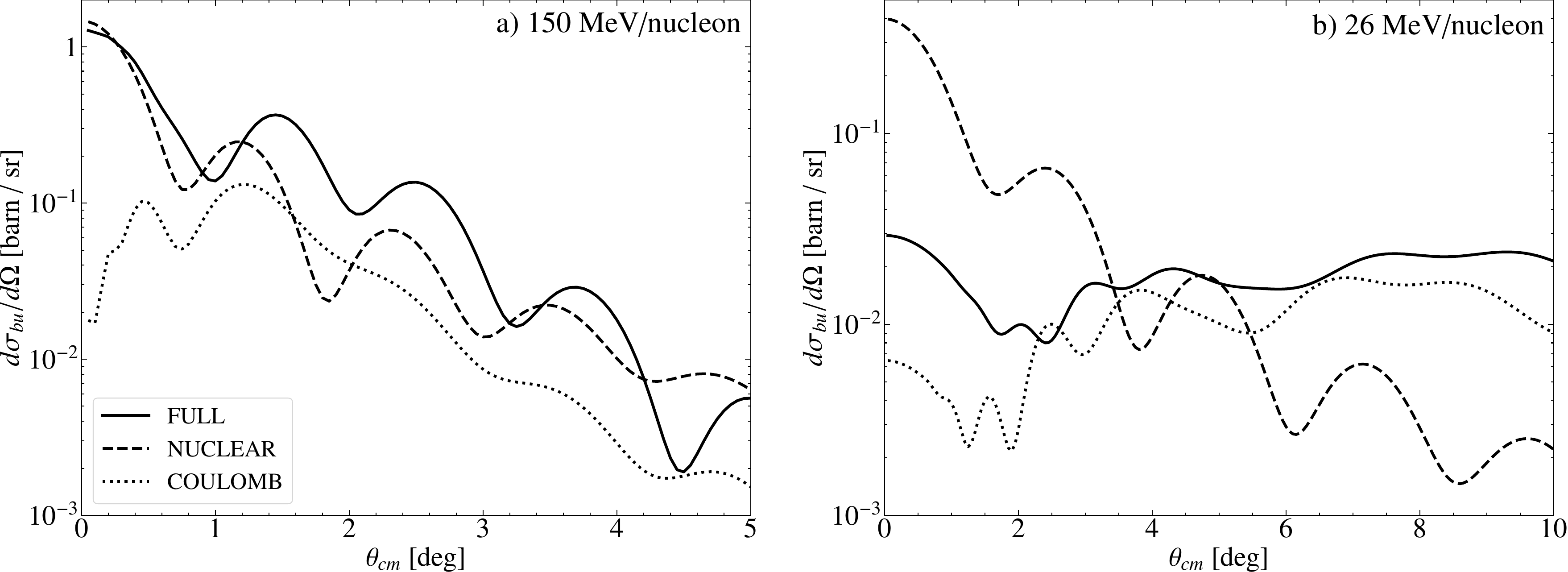}
    \caption{Breakup cross section as a function of the scattering angle integrated over the energy range $E_{\alpha d}$ = $0$--$500 $ keV at a) 150 MeV/nucleon and b) 26 MeV/nucleon.}
    \label{fig:angular}
\end{figure*}

\section{Breakup-reaction analysis}

Within the DEA, we perform a theoretical analysis of the Coulomb dissociation of $\prescript{6}{}{\rm Li}$ on $\prescript{208}{}{\rm Pb}$ experimentally measured at 150 MeV/nucleon by \citet{Hammache}, and at 26 MeV/nucleon by \citet{Kiener}. In each analysis shown here, $\prescript{6}{}{\rm Li}$ is described with the inter-fragment potential $V_{\alpha d}$ developed in Ref.~\cite{Hammache}, which leads to an excellent reproduction of the bound and scattering properties of the $\alpha$-$d$ system.

For the breakup reaction at 150 MeV/nucleon the cluster-target optical potentials were also taken from Ref. \cite{Hammache}, where they were generated from systematic optical potential models for nucleon-Pb scattering by a double-folding technique. For the breakup at 26 MeV/nucleon, the $\alpha$-$T$ and $d$-$T$ optical potentials were retrieved from the global parametrizations of \citet{Nolte} and \citet{An}, respectively.

Our results are shown in Figs. \ref{fig:energy} and \ref{fig:angular}. To analyze the role of the different interactions in the breakup process, three different calculations were performed in each case: a full calculation, in which the $\prescript{6}{}{\rm Li}$ projectile and the $\prescript{208}{}{\rm Pb}$ target interact via both Coulomb and nuclear forces (solid lines), and also pure Coulomb (dotted lines) and pure nuclear (dashed lines) calculations, in which the projectile-target system interacts via only one particular interaction.  In the case of the reaction involving only Coulomb interaction, a cut is imposed on the impact parameter to simulate absorption by the nuclear target.

Figure \ref{fig:energy}a) presents the energy distribution of the breakup cross section integrated over all angles for the reaction at 150 MeV/nucleon measured by \citet{Hammache}. The dash-dotted line is the incoherent sum of pure Coulomb and pure nuclear calculations. In the energy range of BBN, shown by the yellow bar, pure nuclear breakup dominates the reaction. Moreover, the differences between the full calculation and the incoherent Coulomb plus nuclear sum highlight the presence of Coulomb-nuclear interferences at all relative energies.

Figure \ref{fig:angular}a) presents the corresponding angular distribution of the breakup cross section integrated over the domain of relative energies relevant for BBN, that is, $E_{\alpha d} \in [0.0, 0.5]$ MeV. The angular distribution of the full breakup process is heavily influenced by the nuclear contribution in the whole angular range, and there are evident Coulomb-nuclear interferences, as manifested already in the energy distribution.

Our results confirm the conclusions drawn in the analysis of \citet{Hammache} suggesting that the breakup process at 150 MeV/nucleon is nuclear dominated in the energy range of interest for BBN and involves Coulomb-nuclear interferences.

The energy distribution from the breakup at 26 MeV/nucleon is shown in Fig.~\ref{fig:energy}b). In the energy range of BBN, the magnitude of pure Coulomb and pure nuclear breakup processes are comparable to each other. Moreover, the presence of interferences remains at relative energies above 0.4 MeV. The angular distribution in this case is shown in Figure \ref{fig:angular}b). Here, the Coulomb interaction plays a stronger role in the reaction. Nevertheless, there is no consistent agreement between the full breakup cross section and the pure Coulomb one.

The results for the angular distributions in Fig. \ref{fig:angular} show that the dissociation of $\prescript{6}{}{\rm Li}$ by the interaction with $\prescript{208}{}{\rm Pb}$ is nuclear dominated at forward angles because the Coulomb contribution is, at least, several times smaller in magnitude. We establish that such a forward suppression of pure Coulomb breakup is a consequence of the $\alpha$-$d$ cluster structure of $\prescript{6}{}{\rm Li}$. In a semiclassical view of the reaction, forward-angle contributions result from interactions at large impact parameters, where the Coulomb field dominates. In this particular case, the $\alpha$ and deuteron fragments in the projectile are both $N=Z$ nuclei and, therefore, possess approximately the same charge-to-mass ratio. Due to this symmetry, both fragments are equally accelerated at large impact parameters by the Coulomb field of the target resulting in no tidal force to induce the projectile's dissociation \cite{Baye_2018}. This imprint of the nuclear structure onto the reaction dynamics is what leads to the reduction of Coulomb dissociation. This particular observation has broader implications in the application of the Coulomb breakup method for the study of astrophysical reactions involving other $N=Z$ clustered nuclei such as $\prescript{16}{}{\rm O}$ \cite{Bott:2024hxh, bertulani1994}.

The generality of these results and qualitative conclusions has been verified against different fragment-target optical potentials as well as different two-cluster descriptions of $\prescript{6}{}{\rm Li}$ \cite{Hammache, tursunov2015theoretical, dubovichenko}, including new effective cluster potentials based on the ab-initio description of Ref.~\cite{hebborn} that will be presented in a work in preparation \cite{sanjinez}.

\begin{figure}[!t]
\centering
    \includegraphics[width=0.92\linewidth]{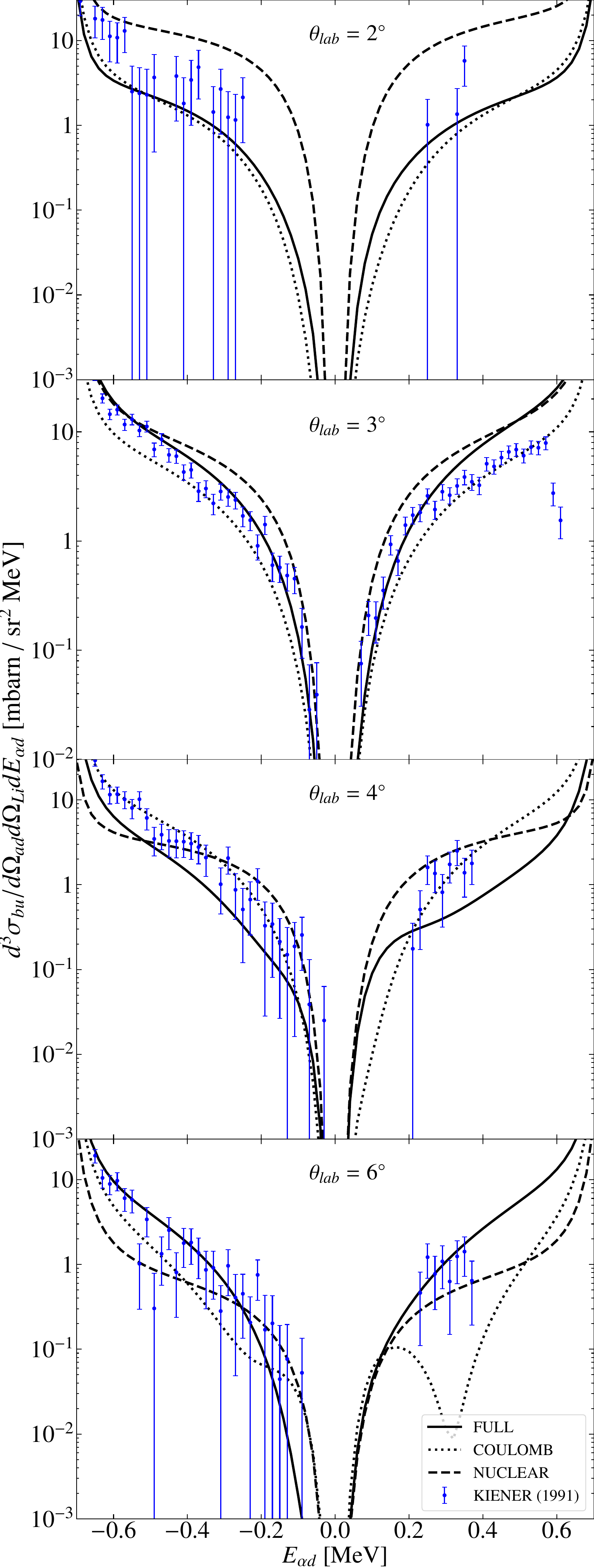}
    \caption{Triple differential breakup cross section at 26 MeV/nucleon compared to the data measured by \citet{Kiener}. The positive and negative $E_{\alpha d}$ branches correspond, respectively, to the forward and backward emission of the $\alpha$ in the reaction.}
     \label{fig:triple}
\end{figure}

The experimental campaign of \citet{Kiener} selected breakup cross section measurements at forward angles under the semiclassical assumption that Coulomb breakup dominates at angles below the rainbow condition $\theta_{lab} = 7 \degree$. Therefore, they collected data for the triple differential breakup cross section $d^3 \sigma / d\Omega_{\alpha} d \Omega_d dE_{\alpha d}$ at different $\theta_{lab}$ and converted the measurements to the center-of-mass frame cross section $d^3 \sigma / d\Omega_{\alpha d} d \Omega_{Li} dE_{\alpha d}$ to infer the astrophysical S-factor.

Figure \ref{fig:triple} shows our theoretical calculations for the triple differential cross section $d^3 \sigma / d\Omega_{\alpha d} d \Omega_{Li} dE_{\alpha d}$ along with the experimental data from \citet{Kiener} at $\theta_{lab} = 2 \degree, 3 \degree, 4 \degree$, and $6 \degree$ transformed according to Refs. \cite{fuchs1982cross, Kiener2}. Our full calculation compares very well with the measurements, especially at $\theta_{lab} = 3 \degree$, which have been used by \citet{Kiener} to infer the astrophysical S-factor for the $\alpha(d,\gamma)\prescript{6}{}{\rm Li}$ capture reaction. Note that \citet{Kiener} refers to issues in the acceptance of their spectrograph that limited their measurements in the positive branch of $E_{\alpha d}$. The effects of this can be observed around and beyond 0.6 MeV in the plotted data.

The agreement of our calculations with experiment shows that nuclear contributions play an important role at every angle and relative energy. It also demonstrates the strong role of Coulomb-nuclear interferences. In addition, it highlights the validity of using few-body models for the study of low-energy nuclear reactions, which are characterized by their peripheral nature, as intended in Ref.~\cite{Hammache}.

\section{On the inference of the S-factor in \citet{Kiener}}

The analysis of \citet{Kiener} determined an approximately constant S-factor in the energy range of interest for BBN, a result that has been in tension with several theoretical predictions \cite{Hammache, hebborn} and with the direct measurements of the LUNA collaboration \cite{LUNA}. However, their work was based on some important hypotheses. First, they assumed that Coulomb breakup dominates at forward angles. Thus, they neglected the nuclear contribution and the presence of interferences in their analysis. Second, they assumed that the dissociation process is dominated by a pure electric quadrupole transition (E2) and, therefore, that the radiative capture cross section can be inferred considering only outgoing $d$-wave contributions starting from the $s$-wave bound state of $\prescript{6}{}{\rm Li}$.

As discussed in the previous section, our results for the angular distribution and the triple differential breakup cross section indicate that the first assumption does not hold for this reaction. We have shown that nuclear breakup dominates at forward angles due to the $N=Z$ cluster structure of the $\prescript{6}{}{\rm Li}$ projectile and that the measurements at $\theta_{lab} = 3 \degree$ are consistent with a breakup process that includes nuclear contributions. Therefore, our results provide evidence that the S-factor inferred by \citet{Kiener} was partially enhanced by unaccounted additional contributions from nuclear breakup.

\begin{figure}
\begin{center}
    \includegraphics[width=0.92\linewidth]{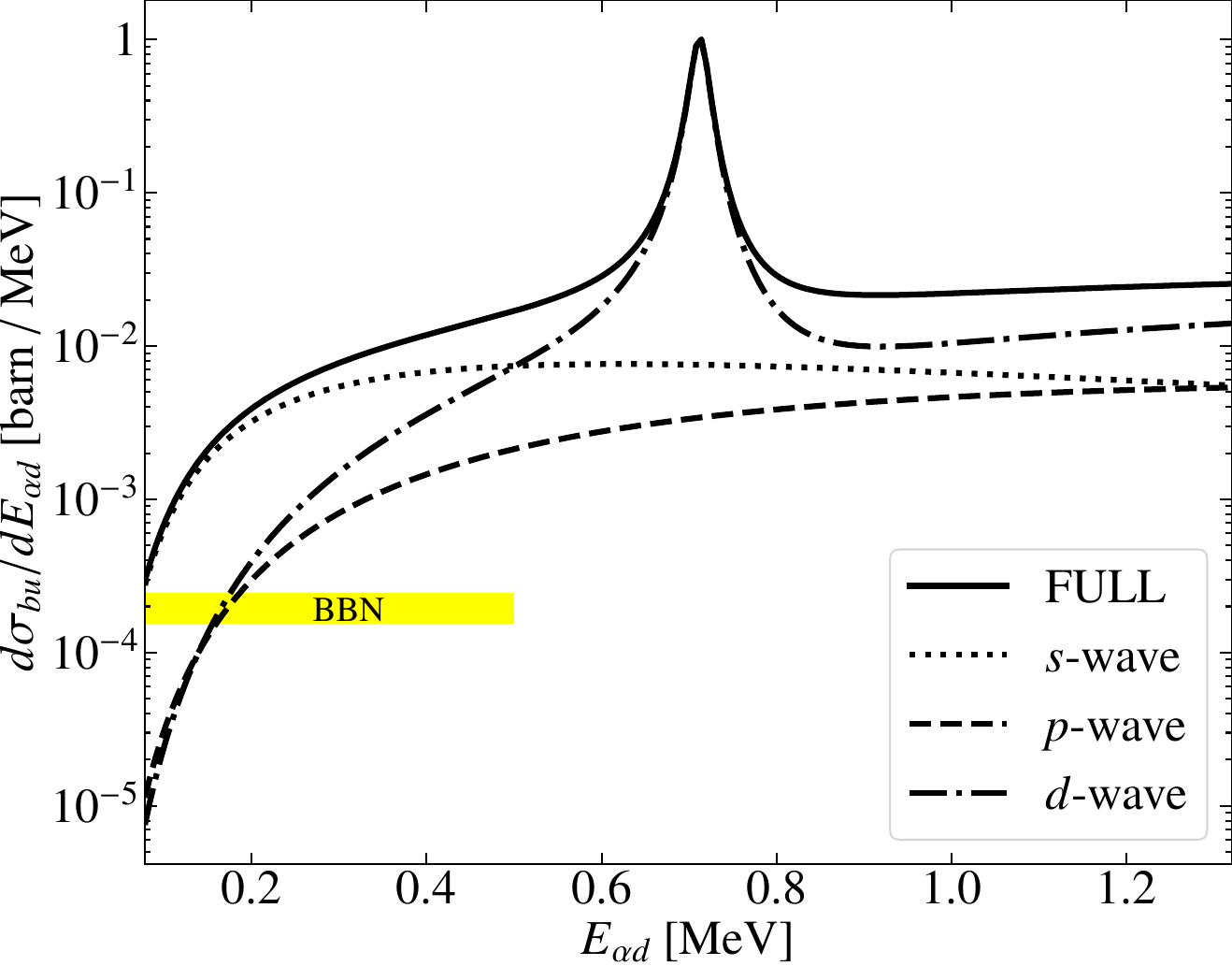}
    \caption{Partial-wave decomposition of the full breakup cross section at 26 MeV/nucleon.}
     \label{fig:partial}
\end{center}
\end{figure}

Moreover, our results demonstrate that the second assumption regarding the dominance of a single electric quadrupole in the process does not hold either. Figure \ref{fig:partial} shows the partial-wave decomposition of the full breakup cross section at 26 MeV/nucleon, in which it is evident that $s$- and $p$-wave contributions play also a relevant role at energies of astrophysical interest. The same observation holds for the breakup reaction at 150 MeV/nucleon. The $s$-wave contribution, in particular, dominates over the entire energy range of interest for BBN and could lead to a significant enhancement of the inferred S-factor. This contribution arises in the dynamical reaction model from nuclear-induced breakup, and from higher-order corrections to the Coulomb process \cite{Bertsch, coupling} that can be connected to multi-photon exchange effects that have been observed and studied in detail in Ref. \cite{Typel}.

\section{Summary and conclusions}

In this work, we present the results of a theoretical re-analysis of the Coulomb dissociation of $\prescript{6}{}{\rm Li}$ on $\prescript{208}{}{\rm Pb}$ experimentally determined at 26 MeV/nucleon by \citet{Kiener} and at 150 MeV/nucleon by \citet{Hammache}. Our findings confirm that at both beam energies the breakup of $\prescript{6}{}{\rm Li}$ is heavily influenced by the nuclear interaction with the $\prescript{208}{}{\rm Pb}$ target.

We have found that, contrary to semiclassical expectations, this breakup reaction is generally nuclear dominated at forward angles due to a suppression of the Coulomb contribution. This suppression results from the $N=Z$ cluster structure of $\prescript{6}{}{\rm Li}$, due to which the fragments in the projectile are equally accelerated by the Coulomb field at large impact parameters. The breakup of the projectile is thus significantly reduced. We assert that this particular finding may have broader implications in the application of the Coulomb breakup method for the study of astrophysical reactions involving other $N=Z$ clustered nuclei, such as $\prescript{16}{}{\rm O}$ \cite{Bott:2024hxh, bertulani1994}.

Our comparison with the triple differential breakup cross section measured by \citet{Kiener} at $\theta_{lab} = 3 \degree$ indicates that their S-factor was inferred from data that included nuclear breakup contributions and higher-order effects that were unaccounted for within their model of the reaction. This certainly affected the inference of the astrophysical S-factor from breakup data. 

These observations render the determination of the radiative capture cross section from breakup reaction data in the case of $N=Z$ clustered nuclei extremely challenging. Thus, we underscore the importance of advancing alternative indirect methods for the study of reactions of astrophysical interest such as electron-induced dissociation.

\section*{Declaration of competing interest}

The authors declare that they have no knowledge of competing financial or personal interests that could have influenced the work reported in this article.

\section*{Acknowledgments}
This project was funded by the Deutsche Forschungsgemeinschaft (DFG, German Research Foundation) – project number: 514321794 (CRC1660: Hadron and Nuclei as discovery tools) and the PRISMA++ (Precision Physics, Fundamental Interactions and Structure of Matter) Cluster of Excellence.

\appendix

\bibliographystyle{unsrtnat}
\bibliography{references}

\end{document}